\documentstyle[12pt]{article}
\makeatletter
 
\def\section{\@startsection {section}{1}{\z@}{+3.0ex plus +1ex minus
  +.2ex}{2.3ex plus .2ex}{\normalsize\bf}}
 
\expandafter\ifx\csname mathrm\endcsname\relax\def\mathrm#1{{\rm #1}}\fi
 
\makeatletter

\newcount\@tempcntc
\def\@citex[#1]#2{\if@filesw\immediate\write\@auxout{\string\citation{#2}}\fi
  \@tempcnta\z@\@tempcntb\m@ne\def\@citea{}\@cite{\@for\@citeb:=#2\do
    {\@ifundefined
       {b@\@citeb}{\@citeo\@tempcntb\m@ne\@citea
        \def\@citea{,\penalty\@m\ }{\bf ?}\@warning
       {Citation `\@citeb' on page \thepage \space undefined}}%
    {\setbox\z@\hbox{\global\@tempcntc0\csname
b@\@citeb\endcsname\relax}%
     \ifnum\@tempcntc=\z@ \@citeo\@tempcntb\m@ne
       \@citea\def\@citea{,\penalty\@m}
       \hbox{\csname b@\@citeb\endcsname}%
     \else
      \advance\@tempcntb\@ne
      \ifnum\@tempcntb=\@tempcntc
      \else\advance\@tempcntb\m@ne\@citeo
      \@tempcnta\@tempcntc\@tempcntb\@tempcntc\fi\fi}}\@citeo}{#1}}

\def\@citeo{\ifnum\@tempcnta>\@tempcntb\else\@citea
  \def\@citea{,\penalty\@m}%
  \ifnum\@tempcnta=\@tempcntb\the\@tempcnta\else
   {\advance\@tempcnta\@ne\ifnum\@tempcnta=\@tempcntb \else
\def\@citea{--}\fi
    \advance\@tempcnta\m@ne\the\@tempcnta\@citea\the\@tempcntb}\fi\fi}

\def\beq{\begin{equation}}
\def\eeq{\end{equation}}
\def\beqar{\begin{eqnarray}}
\def\eeqar{\end{eqnarray}}
\def\barr#1{\begin{array}{#1}}
\def\earr{\end{array}}
\def\bfi{\begin{figure}}
\def\efi{\end{figure}}
\def\btab{\begin{table}}
\def\etab{\end{table}}
\def\bce{\begin{center}}
\def\ece{\end{center}}

\def\text{\textstyle}

\def\mathswitchr#1{\relax\ifmmode{\mathrm{#1}}\else$\mathrm{#1}$\fi}

\def\mathswitch#1{\relax\ifmmode#1\else$#1$\fi}

\newcommand{\mpar}[1]{{\marginpar{\hbadness10000%
                      \sloppy\hfuzz10pt\boldmath\bf#1}}%
                      \typeout{marginpar: #1}\ignorespaces}
\def\draftdate{\relax}
\def\mda{\relax}
\def\mua{\relax}
\def\mla{\relax}
\def\draft{
\def\thtystars{******************************}
\def\sixtystars{\thtystars\thtystars}
\typeout{}
\typeout{\sixtystars**}
\typeout{* Draft mode!
         For final version remove \protect\draft\space in source file *}
\typeout{\sixtystars**}
\typeout{}
\def\draftdate{\today}
\def\mda{\mpar{$\downarrow$}}
\def\mua{\mpar{$\uparrow$}}
\def\mla{\marginpar[\boldmath\hfil$\rightarrow$\hfil]%
                   {\boldmath\hfil$\leftarrow $\hfil}%
                    \typeout{marginpar:
$\leftrightarrow$}\ignorespaces}
\def\mua{\marginpar[\boldmath\hfil$\uparrow$]%
                   {\boldmath$\uparrow$\hfil}%
                    \typeout{marginpar: $\uparrow$}\ignorespaces}
\def\mda{\marginpar[\boldmath\hfil$\downarrow$]%
                   {\boldmath$\downarrow$\hfil}%
                    \typeout{marginpar: $\downarrow$}\ignorespaces}
\def\mla{\marginpar[\boldmath\hfil$\rightarrow$]%
                   {\boldmath$\leftarrow $\hfil}%
                    \typeout{marginpar: $\leftrightarrow$}\ignorespaces}
\overfullrule 5pt
\oddsidemargin -15mm
\marginparwidth 29mm
}
 
\def\stars{\strut\leaders\hbox{*}\hfill\strut}
\def\starline{\hfil\strut\hfil\hbox to \textwidth {\stars}\hfil}

\begin{document}
 
\begin{titlepage}
 
\title{Implications of Electroweak Precision Tests\footnote{Invited Talk presented at 
the First Arctic Workshop on Particle Physics, Saariselkae, August 1994 and at the 
Workshop on Elementary Particles and Quantum Field Theory, Moscow, September 1994} \footnote{
Supported by Bundesministerium f\"ur Forschung und Technologie (BMFT), Germany}
}
 
\author{
Dieter Schildknecht  \\
Fakult\"at f\"ur Physik, Universit\"at Bielefeld \\
Postfach 10 01 31, D-33501 Bielefeld, Germany
}
 
\date{}
 
\maketitle
\end{titlepage} 

\newpage

\begin{quote}
{\bf
'' In any event, it is always a good idea to try to see how much 
or how little
of our theoretical knowledge actually goes into the 
ana\-lysis of those situations 
which have been experimentally checked.''

\hfill
R.P. Feynman} \cite{Feyn} 
\end{quote}

\vspace{0.3cm}\noindent
The spirit of this talk is best charcterized by the above quotation.
 
\vspace{1cm}\noindent
{\bf Tree-level Predictions}
 
\vspace{0.3cm}
In this spirit, let us look at the implications of electroweak precision data
from LEP and the $W^\pm$ mass. The quality of these data is best appreciated by
starting from the tree-level predictions. From the input of 
\begin{eqnarray*}
\alpha (0)^{-1} & = & 137.0359895(61) \\
G_\mu & = & 1.16639(2) \cdot 10^{-5} {\rm GeV}^{-2} \\
M_Z & = & 91.1899 \pm 0.0044 {\rm GeV}
\end{eqnarray*}
one may predict the partial width of the $Z^0$ for decay into leptons, $\Gamma_l$,
the weak mixing angle, $\bar s^2_W$, and the mass ratio, $M_{W^\pm}/ M_Z$. 
A comparison of these data with the tree-level predictions shows that
the simple $\alpha (0)$ tree-level prediction fails by several standard deviations.
The $\alpha (0)$ tree-level prediction yields, 
\begin{eqnarray*}
\bar s^2_W (th) & = & 0.2121, \\
\Gamma_l (th) &=& 84.85~{\rm MeV}, \\
\frac{M_{W^\pm} (th)}{M_Z} &=& 0.8876 ,
\end{eqnarray*}
which is to be compared with
the experimental data \cite{Mor,Glas} 
\begin{eqnarray*}
\bar s^2_W ({\rm all~asymm.~LEP}) & = 0.23223 \pm 0.00050  \\
\bar s^2_W ({\rm all~asymm.~LEP~+~SLD}) & = 0.23158 \pm 0.00045 \\
\Gamma_l & = 83.98 \pm 0.18~{\rm MeV} \\
{{M_{W^\pm}}\over{M_{Z^0}}} & = 0.8814 \pm 0.0021
\end{eqnarray*}
  
\vspace{1cm}\noindent
{\bf Loop-Effects}
 
\vspace{0.3cm}
Concerning loop effects, I follow the 1988 stategy of Gounaris and myself, ''to 
isolate and to test directly the ''new physics'' of boson loops and other new
phenomena by comparing with and looking for deviations from the predictions of the 
dominant-fermion-loop results''\cite{Goun}, i.e., let us 
discriminate between fermion-loop vacuum-polarization contributions to photon
propagation {\it as well as $Z^0$ and $W^\pm$ propagation} on the one hand and 
boson-loop effects on the other hand. The reason for such a distinction is in fact 
obvious: the fermion-loop effects can be precisely predicted from the known 
couplings of the leptons and (light) quarks, while the other loop effects, e.g.
vacuum-polarization involving boson pairs and vertex corrections, depend on 
empirically unknown couplings among the vector bosons (including the Higgs scalar 
boson in the case of bosonic vacuum-polarization diagrams).
In fact, it is the difference between the fermion-loop results and the full 
one-loop results which sets the scale for the precision needed for tests of the 
theory of electroweak interactions beyond (trivial) fermion-loop effects. 
One should remind oneself that the experimentally unknown bosonic interaction 
properties are right at the heart of renormalizability of the electroweak theory. 
The necessary precision for such tests of the theory beyond the leading fermionic
contributions has only been reached by the data presented this year
(Moriond '94 \cite{Mor} and Glasgow conference \cite{Glas})
 
In our analysis \cite{Ditt}, we restrict ourselves to the leptonic observables. 
The extension to hadronic decays is formulated in \cite{Ditt-2}.

In figs. 1 to 3 from \cite{Ditt}, 
we show the above-mentioned experimental data compared with various
theoretical predictions:
 
\begin{itemize}
\item[i)]
The $\alpha (M^2_Z)$ tree-level prediction, which is obtained by taking into 
account the change in the electromagnetic coupling due to leptons and quarks between
the low energy scale of $\alpha (0)$ and the scale 
$M^2_Z$ by the replacement \cite{Jeg}
$$
\alpha (0)^{-1} \rightarrow \alpha (M^2_Z)^{-1} = 128.87 \pm 0.12
$$
in the tree-level formulae. It is represented by the isolated point in 
figs. 1 to 3.

\item[ii)]
The fermion-loop prediction, which takes into account the quark- and 
lepton-loop contributions not only to the photon propagator but also to the 
$Z^0$ and the $W^\pm$ propagator (the latter one entering 
the theoretical predictions via $G_\mu$ and the 
top-quark loop). In figs. 1 to 3 the result is indicated by the lines with square
insertions, marking the assumed mass of the top quark. 
 
\item[iii)] The full one-loop standard model result, which includes all effects due
to vacuum polarization, vertex- and box contributions and consequently depends
on trilinear and quadrilinear couplings of the bosons among each other and the 
mass, $m_H$, of the Higgs boson. 
 
\end{itemize}
 
We conclude that
\begin{itemize}
\item
contributions beyond the $\alpha (M^2_Z)$ tree-level prediction, i.e., 
electroweak corrections (in addition to the purely electromagnetic ones entering
the running of $\alpha (0)$ to $\alpha (M^2_Z)$) are surely needed (a point also
stressed by Okun and collaborators \cite{OK}),
 
\item
contributions beyond {\it the full fermion-loop} results are necessary, 

\item
there is agreement with the full one-loop result of the $SU(2) \times U(1)$ 
theory which provides bosonic loop corrections in addition to the fermion loops. 
\end{itemize}
 
The question immediately arises what can be said about the nature of the 
bosonic loops which lead to the final agreement between theory and 
experiment in figs. 1 to 3.
 
\vspace{1cm}\noindent
{\bf Effective Lagrangian, $\Delta x, \Delta y, \epsilon$ Parameters}
 
\vspace{0.3cm}
This question can best be answered by an analysis in terms of the parameters 
$\Delta x, \Delta y$ and $\epsilon$ which within the framework of an effective 
Lagrangian \cite{Ditt} specify various possible sources of $SU(2)$ violation. 
The parameter $x$ is related to $SU(2)$ violation in the triplet of charged and 
neutral (unmixed) vector boson via
$$
M^2_{W^\pm} = (1+ \Delta x) M^2_{W^0} \equiv xM^2_{W^0},
$$
while $\Delta y$ specifies $SU(2)$ violation among the $W^\pm$ and $W^0$ couplings 
to fermions,
$$
g^2_{W^\pm} (0) \equiv M^2_{W^\pm} 4 \sqrt 2 G_\mu = (1 + \Delta y) g^2_{W^0}
(M^2_Z) \equiv y g^2_{W^0}.
$$
Finally, the parameter $\epsilon$ refers to a mixing strength, when formulating
the theory in terms of current mixing \'a la Hung Sakurai \cite{HS},
$$
{\cal L}_{{\rm mix}} \equiv {{e(M^2_Z)}\over{g_{W^0}(M^2_Z)}} (1 - \epsilon ) A_
{\mu\nu} W^{\mu\nu}_0 .
$$
Describing electroweak interactions of leptons at the $Z^0$ in terms of the 
mentioned effective Lagrangian incorporating the three possible sources of 
$SU(2)$ violation, one predicts for the observables $\bar s^2_W , M_{W^\pm}$ and
$\Gamma_l$,
\begin{eqnarray*}
\bar s^2_W (1 - \bar s^2_W ) & =& {{\pi \alpha (M^2)}\over{\sqrt 2 G_\mu M^2_Z}}
{y\over x} (1 - \epsilon) {1\over{\left( 1 + {\bar s^2_W\over{1 - \bar s^2_W}}
\epsilon \right)}},  \\
{M^2_{W^\pm}\over M^2_Z} & =& (1 - \bar s^2_W ) x \left( 1 + {\bar s^2_W \over
{1 - \bar s^2_W}} \epsilon \right) , \\
\Gamma_l & =& {{G_\mu M^3_Z}\over{24\pi \sqrt 2}} \left( 1 + (1 - 4 \bar s^2_W )^2
\right) {x\over y} \left( 1 - {{3\alpha}\over{4\pi}} \right).
\end{eqnarray*}
For $x = y = 1$ (i.e., $\Delta x = \Delta y = 0$) and $\epsilon = 0$ one
recovers the $\alpha (M^2_Z)$ tree-level results mentioned previously. For later usage,
we introduce the mixing angle $s^2_0$ via
$$
s^2_0 (1 - s^2_0 ) \equiv {{\pi\alpha (M^2_Z)}\over{\sqrt 2 G_\mu M^2_Z}}.
$$
 
By inverting the above relations, $\Delta x , \Delta y$ and $\epsilon$ may 
now be deduced from the experimental data on $\bar s^2_W, \Gamma_l$ and 
$M_{W^\pm}$. On the other hand, $\Delta x , \Delta y$ and $\epsilon$ may be
theoretically determined 
in the standard electroweak theory 
at the one-loop level, always strictly discriminating 
between pure fermion-loop predictions and the rest which contains the unknown 
bosonic couplings. The striking results of such an analysis are shown in figs. 4, 5, 6.
 
According to fig. 4, the data in the $(\epsilon , \Delta x)$ plane are well 
described if $\Delta x$ and $\epsilon$ are approximated by their 
fermion-loop values, 
\begin{eqnarray*}
\Delta x & =& \Delta x_{{\rm ferm}} (\alpha (M^2_Z) , s^2_0 , m^2_t ) + \Delta x_{{\rm
bos}} (\alpha (M^2_Z) , s^2_0 , \ln m^2_H) \\
  & &  \\
  & \cong & \Delta x_{{\rm ferm}} (\alpha (M^2_Z) , s^2_0 , m^2_t) ,  \\
  & &      \\ 
\epsilon & = & \epsilon_{{\rm ferm}} (\alpha (M^2_Z) , s^2_0 , m^2_t ) + 
\epsilon_{{\rm bos}} (\alpha (M^2_Z) , s^2_0 , \ln m^2_H ) \\
  & &  \\
  &\cong & \epsilon_{{\rm ferm}} (\alpha (M^2_Z) , s^2_0 , m^2_t ) .
\end{eqnarray*}
The logarithmic dependence on the Higgs mass, $m_H$, and the small contributions of 
$\Delta x_{{\rm bos}}$ and $\epsilon_{{\rm bos}}$ imply the well-known result 
that the data are very insensitive to the mass of the Higgs scalar. Values between
60 GeV and more than 1 TeV can easily be accomodated \cite{Pas}.
 
In contrast, a striking effect appears in figs. 5 and 6. The theoretical predictions are
clearly inconsistent with the data, unless the fermion-loop contributions to $\Delta y$ 
(denoted by lines with small squares in figs. 4 to 6) are supplemented by an 
additional term, which in the standard electroweak theory contains bosonic effects, 
$$
\Delta y = \Delta y_{{\rm ferm}} (\alpha (M^2_Z) , s^2_0 , \ln m_t ) + \Delta y_{{\rm 
bos}} (\alpha (M^2_Z) , s^2_0). 
$$
Remembering that $\Delta y$ by definition relates the $W^\pm$ coupling measured in 
$\mu^\pm$ decay to the (unmixed) $Z^0$ coupling, 
$$
g^2_{W^\pm} (0) = (1 + \Delta y) g^2_{W^0} (M^2_Z),
$$
it is not surprising that $\Delta y_{{\rm bos}}$ contains vertex and box corrections
originating from $\mu^\pm$ decay as well as vertex corrections at the $Z^0 f\bar f$ vertex.
While $\Delta y_{{\rm bos}}$ obviously depends on the trilinear couplings among the vector
bosons, it is independent of the Higgs mass, $m_H$. (Note that $\Delta y_{{\rm ferm}}$
and $\Delta y_{{\rm bos}}$ are separately unique and gauge-invariant quantities in the 
$SU(2) \times U(1)$ theory.)
 
In conclusion, the experimental data have become accurate enough to be 
{\it sensitive to loop effects which are independent of $m_H$ but depend on the 
self-interactions of the vector bosons, in particular on the trilinear couplings 
entering the $W^\pm f\bar f^\prime$ and $Z^0 f\bar f$ vertex corrections.}
 
\vspace{1cm}\noindent
{\bf Electroweak Interactions in Higgs-less Massive Vector Boson Theory.}
 
\vspace{0.3cm}
As the experimental results for $\Delta x$ and $\epsilon$ are well represented by
neglecting all effects with the exception of fermion loops, and as the bosonic 
contribution to $\Delta y$ which is seen in the data is independent of $m_H$, the 
question as to the role of the Higgs mass and the concept of the Higgs mechanism with 
respect to precision tests immediately arises.
 
More specifically, one may ask the question whether the experimental results, i.e.
$\Delta x, \Delta y , \epsilon$, can be predicted even without the very concept of the 
Higgs mechanism. 
 
In \cite{GRKN} we start from the well-known fact that the standard electroweak 
theory without Higgs particle can credibly be reconstructed within the framework of a 
massive vector-boson theory with the most general mass mixing term which preserves 
electromagnetic gauge invariance. This theory is then cast into a form which is 
invariant under local $SU(2) \times U(1)$ transformations by introducing three auxiliary
scalar fields \'a la Stueckelberg. As a consequence, loop calculations may be carried out
in an arbitrary $R_\xi$ gauge. 
 
Explicit loop calculations show that indeed the Higgs-less observable $\Delta y$, 
evaluated in the massive vector-boson theory (MVB), coincides with $\Delta y$ evaluated in 
the standard electroweak theory, i.e. in particular for the bosonic part,
we have\footnote{Actually, in the standard theory there is an additional term which 
depends on the Higgs mass like $1/m^2_H$ and is irrelevant numerically for 
$m_H \geq 130$ GeV.}
$$
\Delta y_{{\rm bos}}^{{\rm MVB}} \equiv \Delta y_{{\rm bos}}^{{\rm St.M.}}.
$$
As for $\Delta x_{{\rm bos}}$ and $\epsilon_{{\rm bos}}$, one finds that the 
massive-vector-boson theory and the standard model differ by the replacement
$\ln m_H \Leftrightarrow \ln \Lambda $ ,
where $\Lambda$ denotes an ultraviolet cut-off.
For $\Lambda \leq 1$ TeV, accordingly, 
\begin{eqnarray*}
\Delta x^{{\rm MVB}} & \cong \Delta x_{{\rm ferm}}^{{\rm MVB}} = \Delta x_{{\rm ferm}}^{{\rm 
St.M.}}, \\
\epsilon^{{\rm MVB}} &\cong \epsilon_{{\rm ferm}}^{{\rm MVB}} = \epsilon_{{\rm ferm}}^
{{\rm St.M.}}.
\end{eqnarray*}
 
In conclusion, the massive-vector-boson theory can indeed be evaluated at one-loop level
at the expense of introducing a logarithmic cut-off, $\Lambda$. This cut-off only 
affects $\Delta x$ and $\epsilon$, whose bosonic contributions cannot be resolved 
experimentally. 
 
The quantity $\Delta y$, whose bosonic contributions are essential for agreement with 
experiment, is independent of the Higgs mechanism. It depends on the trilinear couplings
of the vector bosons among each other, which enter the vertex corrections at the 
$W^\pm$ and $Z^0$ vertices. Even though the data cannot discriminate between the massive
vector-boson theory and the standard model with Higgs scalar, the Higgs mechanism 
yields nevertheless the only known simple physical realization of the cut-off $\Lambda$
(by $m_H$) which guarantees renormalizability. 
 
\vspace{1cm}\noindent
{\bf Conclusions}
 
\vspace{0.3cm}
\begin{itemize}
\item
The analysis of the $Z^0$ data and the $W^\pm$ mass in terms of an effective Lagrangian 
with $SU(2)$ beaking via $\Delta x , \Delta y$ and $\epsilon$ yields for these parameters
values which are of the order of magnitude of radiative corrections. This in itself 
consitutes a major triumph of the $SU(2)_L \times U(1)_Y$ symmetry principle which
is at the root of present-day electroweak theory.
 
\item
The data have reached such a high precision that contributions to the parameter 
$\Delta y$ are needed beyond the ones induced by (vacuum polarization) fermion loops to 
the photon, $Z^0$ and $W^\pm$ propagators. These contributions are connected with vertex 
corrections at the $W^\pm f \bar f^\prime$ and $Z^0 f \bar f$ vertices which contain
truely non-Abelian (trilinear) couplings among the vector bosons. 
 
\item
The parameters $\Delta x$ and $\epsilon$, consistently reproduce the data (for $m_t
\simeq 175$ GeV), if approximated by fermion loops, $\Delta x \cong \Delta x_{{\rm ferm}}$
and $\epsilon \approx \epsilon_{{\rm ferm}}$. 
 
\item
The data by themselves do not discriminate a massive-vector-boson theory from the
standard theory based on the Higgs mechanism. The issue of mass generation will remain
open until the Higgs scalar will be found - or something else?
\end{itemize}
 
\vspace{1cm}\noindent
{\bf Acknowledgement}
 
\vspace{0.3cm}
It is a pleasure to thank Misha Bilenky, Stefan Dittmaier, Carsten Gro{\ss}e-Knetter, 
Karol Kolodziej and Masaaki Kuroda for a fruitful collaboration on 
electroweak interactions.

\clearpage

\begin{center}
\begin{picture}(17,25)
\put(0,14.7){\includegraphics{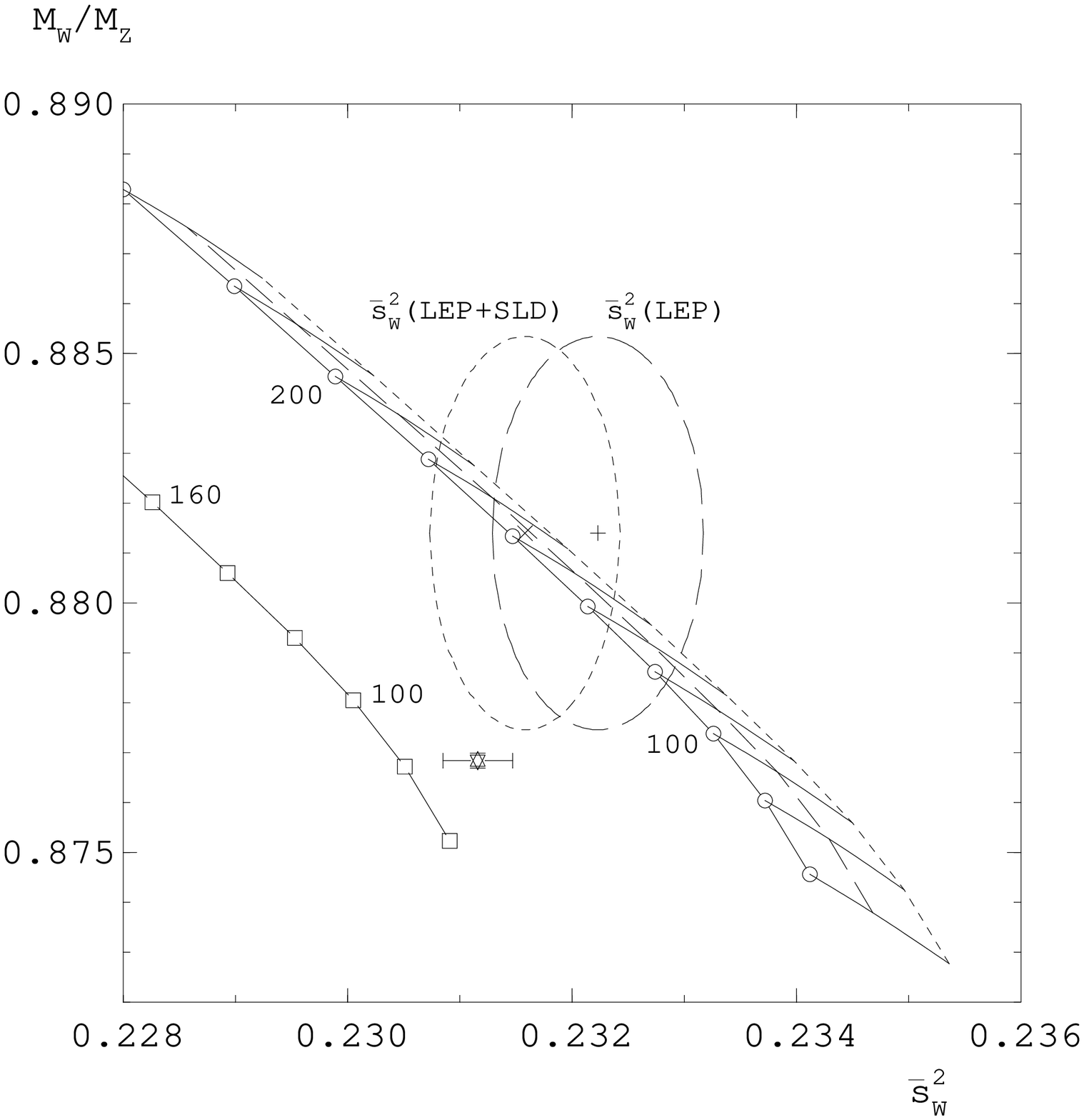}}
\put(0, 6.7){\includegraphics{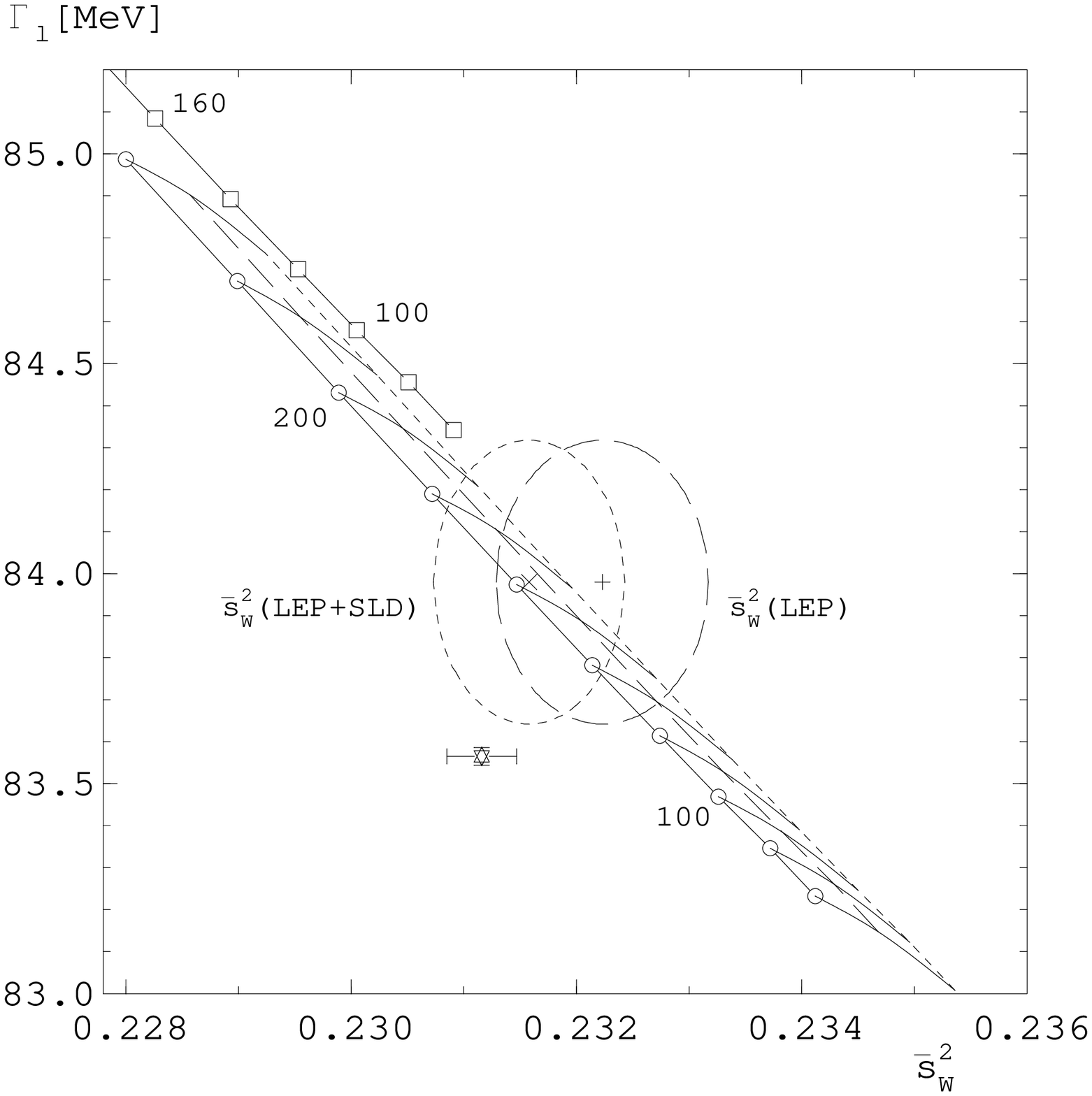}}
\put(0,-1.3){\includegraphics{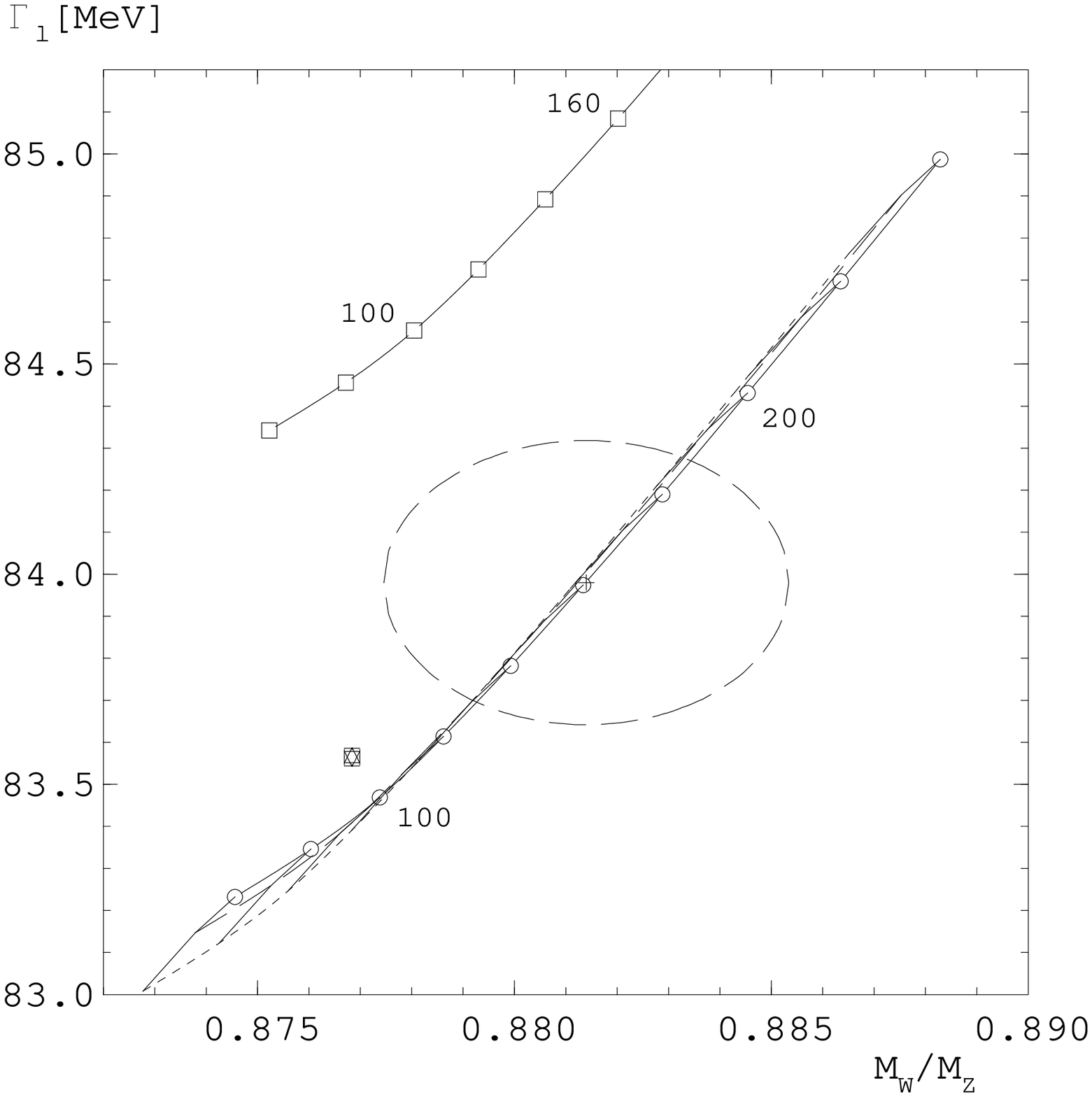}}
\put(9,14.7){\includegraphics{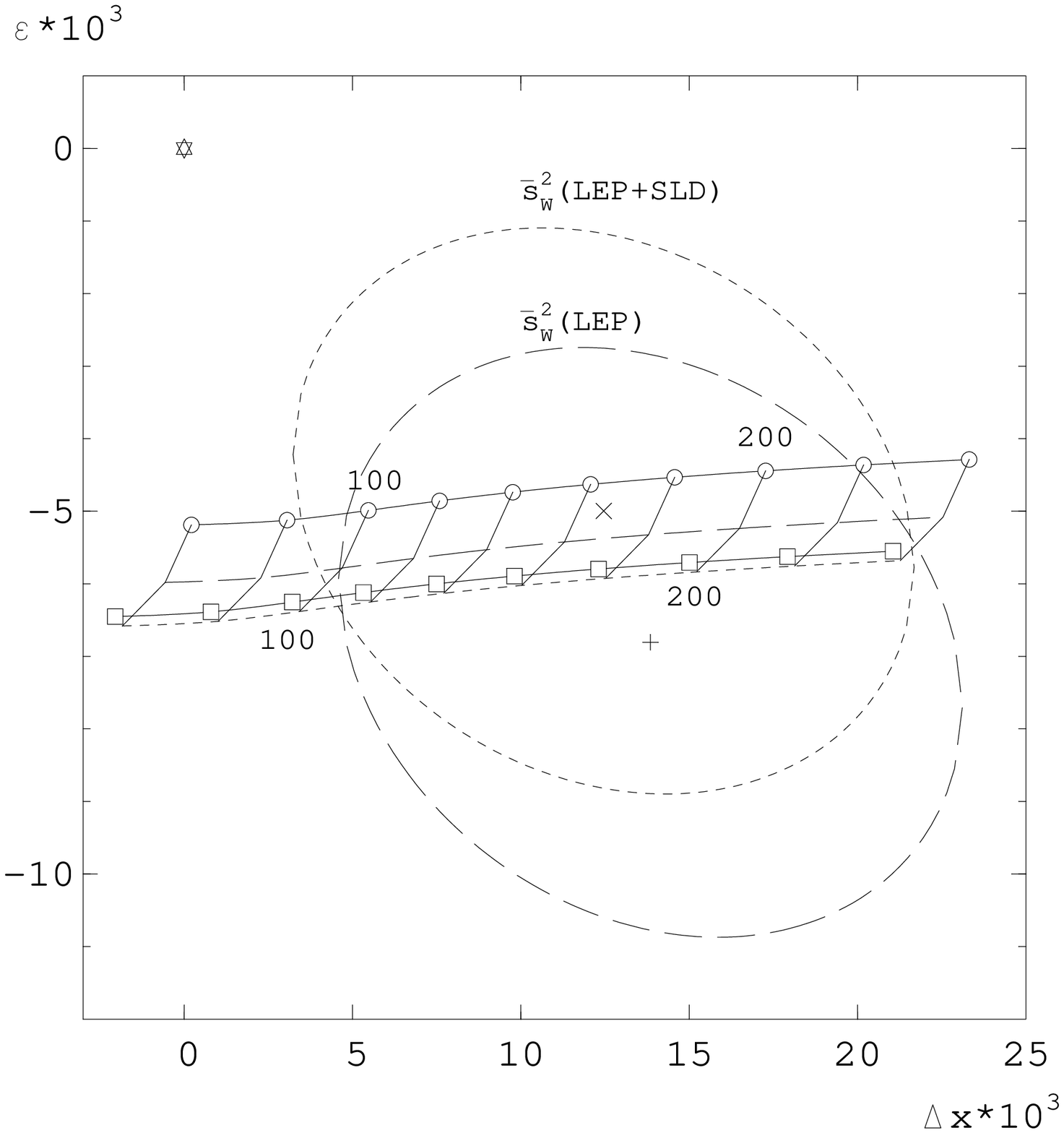}}
\put(9, 6.7){\includegraphics{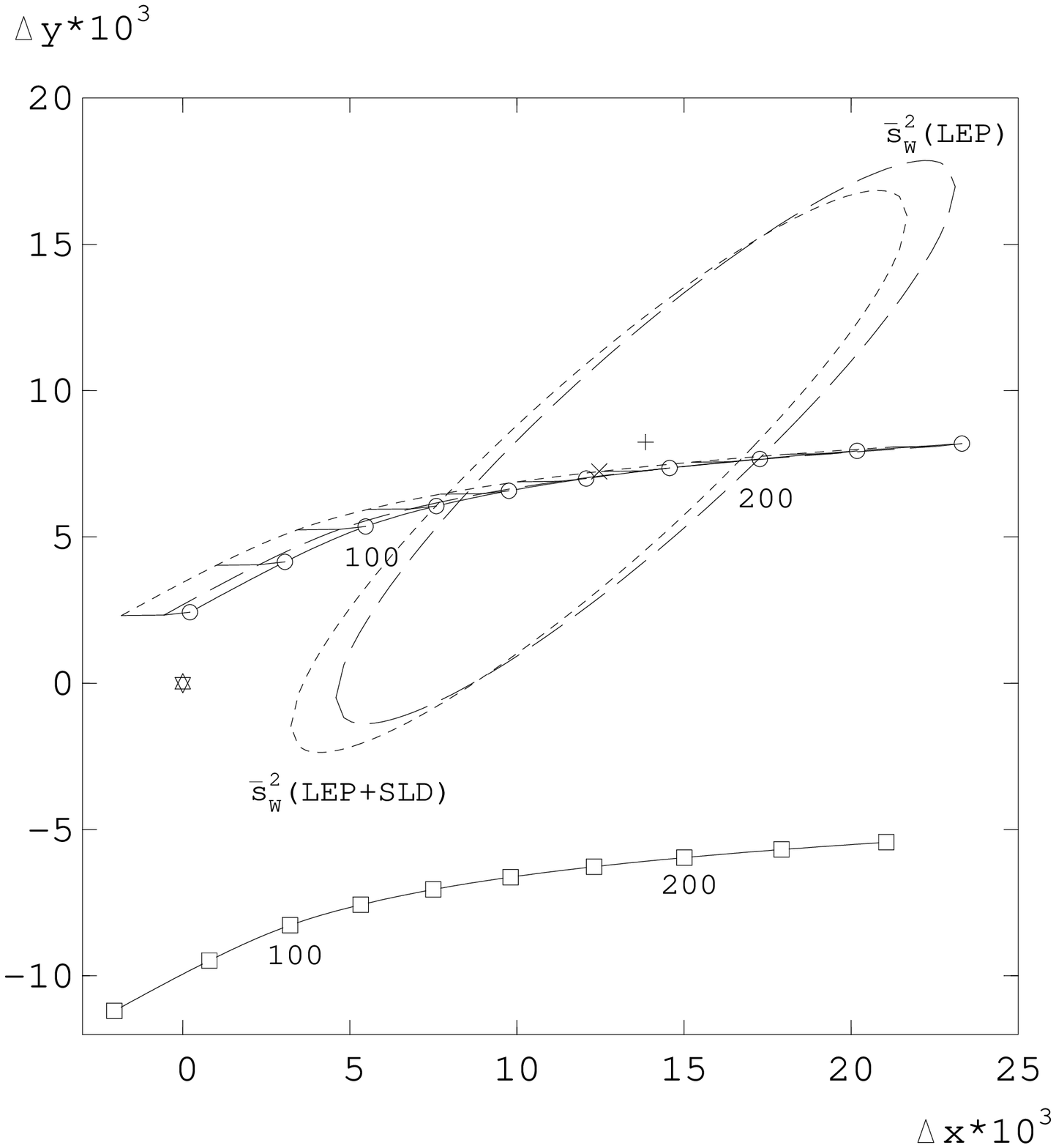}}
\put(9,-1.3){\includegraphics{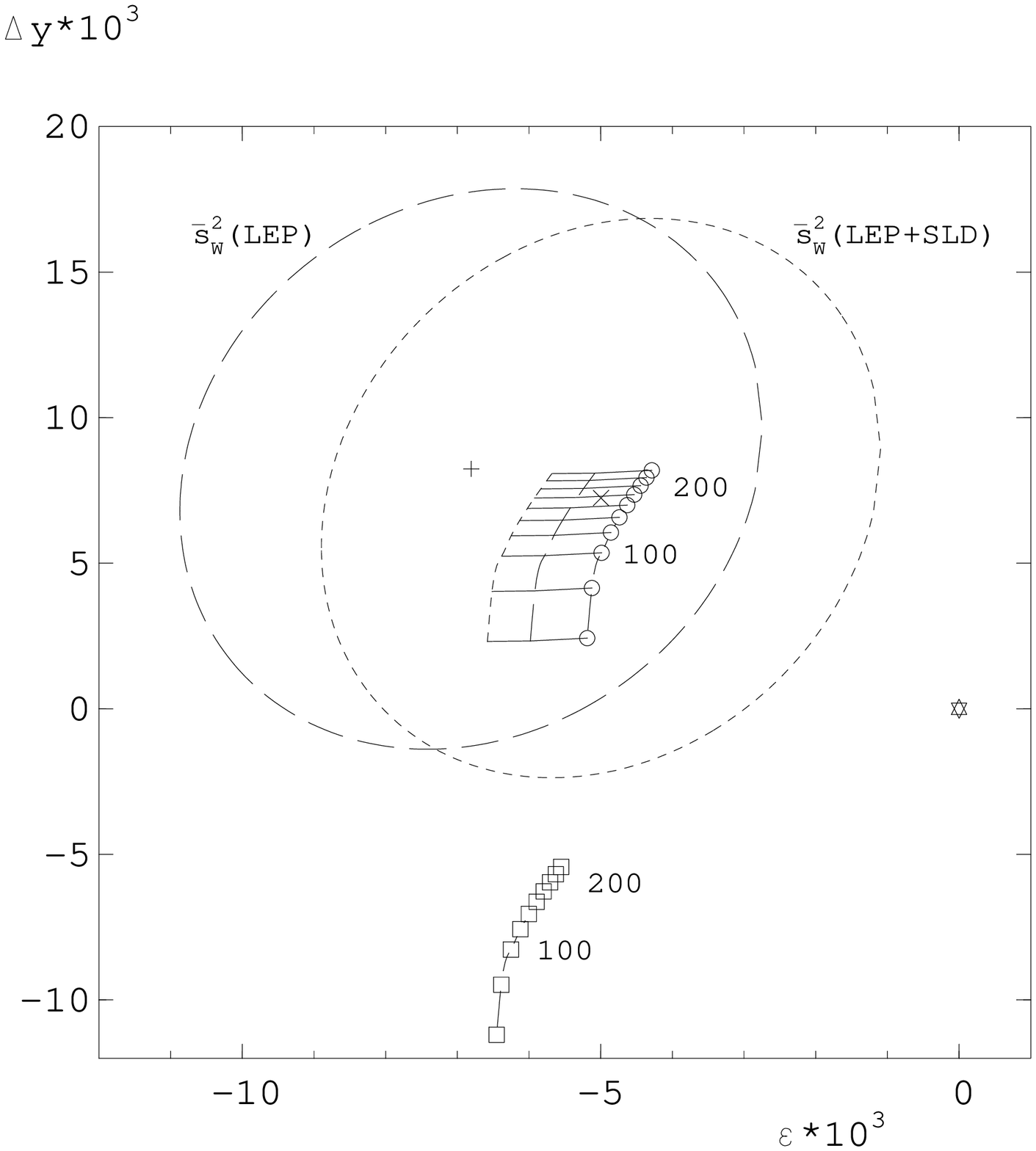}}
\put(3,24.6){Figure 1:}
\put(3,16.6){Figure 2:}
\put(3, 8.6){Figure 3:}
\put(12,24.6){Figure 4:}
\put(12,16.6){Figure 5:}
\put(12, 8.6){Figure 6:}
\put(.8,0.3){\parbox{6.9cm}{
\setlength{\baselineskip}{8pt}
Figs.\ 1,2,3: The experimental data on $(M_W / M_Z ,\, \bar s^2_W ,
\Gamma_l)$ compared with theory.}}
\put(9.8,0.3){\parbox{6.9cm}{
\setlength{\baselineskip}{8pt}
Figs.\ 4,5,6: The experimental data on $\Delta x, \Delta y, \epsilon$
compared with theory.}}
\end{picture}
\end{center}

\end{document}